\documentclass{webofc}
\usepackage[varg]{txfonts}   
\usepackage{wrapfig}
\begin{document}
\title{The Three Hundred--NIKA2 Sunyaev--Zeldovich Large Program twin samples: Synthetic clusters to support real observations}
%
%

\author{\firstname{A.} \lastname{Paliwal}\inst{1}\fnsep \thanks{\email{aishwarya.paliwal@uniroma1.it}} \and
        \firstname{E.} \lastname{Artis}\inst{2} \and
        \firstname{W.} \lastname{Cui}\inst{3} \and
        \firstname{M.} \lastname{De Petris}\inst{1} \and
        \firstname{F.-X.} \lastname{Désert}\inst{4} \and
        \firstname{A.} \lastname{Ferragamo}\inst{1} \and
        \firstname{G.} \lastname{Gianfagna}\inst{1} \and
        \firstname{F.} \lastname{Kéruzoré}\inst{2} \and
        \firstname{J.-F.} \lastname{Macías-Pérez}\inst{2} \and
        \firstname{F.} \lastname{Mayet}\inst{2} \and
        \firstname{M.} \lastname{Muñoz-Echeverría}\inst{2} \and
        \firstname{L.} \lastname{Perotto}\inst{2} \and
        \firstname{E.} \lastname{Rasia}\inst{5,6} \and
        \firstname{F.} \lastname{Ruppin}\inst{7} \and
        \firstname{G.} \lastname{Yepes}\inst{8}
}

\institute{Dipartimento di Fisica, Sapienza Università di Roma, Piazzale Aldo Moro 5, 00185 Roma, Italy
\and
           Univ. Grenoble Alpes, CNRS, Grenoble INP, LPSC-IN2P3, 53, avenue des Martyrs, 38000 Grenoble, France
\and
           Institute for Astronomy, University of Edinburgh, Royal Observatory, Edinburgh EH9 3HJ, United Kingdom
\and
          Univ. Grenoble Alpes, CNRS, IPAG, 38000 Grenoble, France
\and 
          National Institute for Astrophysics, Astronomical Observatory of Trieste (INAF-OATs), via Tiepolo 11, 34131 Trieste, Italy
\and
          Institute for Fundamental Physics of the Universe (IFPU), via Beirut 2, 34014 Trieste, Italy
\and
          Kavli Institute for Astrophysics and Space Research, Massachusetts Institute of Technology, Cambridge, MA 02139, USA
\and
          Departamento de Física Teórica and CIAFF, Módulo 8, Facultad de Ciencias, Universidad Autónoma de Madrid, 28049 Madrid, Spain
          }

\abstract{%
The simulation database of {\sc The Three Hundred} Project has been used to pick synthetic clusters of galaxies with properties close to the observational targets of the NIKA2 camera Sunyaev--Zeldovich (SZ) Large Program. Cross--matching of cluster parameters such as mass and redshift of the cluster in the two databases has been implemented to generate the so--called twin samples for the Large Program. This SZ Large Program is observing a selection of galaxy clusters at intermediate and high redshift $\left( 0.5 < z < 0.9 \right)$, covering one order of magnitude in mass. These are SZ--selected clusters from the Planck and Atacama Cosmology Telescope catalogs, wherein the selection is based on their integrated Compton parameter values, $Y_{500}$: the value of the parameter within the characteristics radius $R_{500}$.

{\sc The Three Hundred} hydrodynamical simulations provide us with hundreds of clusters satisfying these redshift, mass, and $Y_{500}$ requirements. In addition to the standard post-processing analysis of the simulation, mock observational maps are available mimicking X--ray, optical, gravitational lensing, radio, and SZ observations of galaxy clusters. The primary goal of employing the twin samples is to compare different cluster mass proxies from synthetic X--ray, SZ effect and optical maps (via the velocity dispersion of member galaxies and lensing $\kappa$-maps) of the clusters. Eventually, scaling laws between different mass proxies and the cluster mass will be cross--correlated to reduce the scatter on the inferred mass and the mass bias will be related to various physical parameters.
}
\maketitle

\section{Introduction}
\label{intro}
Galaxy clusters are the most massive gravitationally bound objects in the Universe, forming at the nodes of the cosmic web \cite{cweb2,cweb1}. They are excellent tools for both astrophysical and cosmological studies (see \textit{e.g.} \cite{p24,schell,angie}). The imprint of structure formation is carried by the galaxy cluster number density and spatial distribution of galaxy clusters. Hence, the abundance of galaxy clusters with mass and redshift is a well--known cosmological probe. The mass function of galaxy clusters describes the number density of clusters above a given mass $M$. The sensitivity of the cluster distribution, to the underlying cosmology, is well reflected by the mass function of galaxy clusters, which is exponentially sensitive to cosmological parameters \cite{PS}.

Understandably, the cluster mass is a key parameter for studies that aim to constrain cosmological parameters using galaxy cluster distribution with mass and redshift, \textit{i.e.} galaxy cluster number density. The cluster mass, however, is not directly observable and there are several ways to infer it via observations under simple theoretical assumptions. Each of these ways to infer the mass is prone to a bias and intrinsic scatter associated with the observations being used. The cluster mass can be estimated using the hydrostatic equilibrium (HE) equation under the assumption of the cluster having a spherical Dark Matter halo; the Intra Cluster Medium (ICM) profiles, namely, the density profile from X--ray and the pressure profile from the Sunyaev--Zeldovich (SZ) effect \cite{sz} observations, can be combined to infer the cluster mass (see \textit{e.g.} \cite{SZmass}). This inference is prone to mass bias associated to the assumptions made for the HE equation. The kinematic studies of the member galaxies can also be used to estimate the mass of galaxy clusters (see \textit{e.g.} \cite{massdy}). These calculations require spectral information of a lot of member galaxies to get an accurate mass estimate with good precision. Estimating the cluster mass through measuring the gravitational lensing effect of clusters on the background sources is one of the least biased methods but, comes with high dispersion on the mass estimate \cite{wl}. Finally, another way to infer the cluster mass is through scaling laws, which, when correctly calibrated, relate the mass to observational quantities, such as X--ray luminosity, SZ brightness, etc. \cite{p20,schell,esra,aarti,erowl}. An alternative and orthogonal approach to refine the cluster mass estimates is by using state--of--the--art N--body hydrodynamical simulations as useful and efficient test--beds to increase our theoretical understanding \cite{ymmusic}. Evidently, it is critical to understand astrophysical details that may affect the mass estimation via the observational methods mentioned above and to properly calibrate scaling relations between observables and the mass of the cluster.

Establishing the relations between observables and the cluster mass from direct observations is, however, difficult and not a straightforward process. The combination of high sensitivity and high angular resolution observations with data from detailed hydrodynamical simulations of clusters could alleviate these problems. This combination will enable one to accurately estimate the cluster mass and minimise the scatter on the inferred value. This work aims to achieve this goal by the use of observations from the NIKA2 SZ Large Program \cite{lpsz} along with simulated data from {\sc The Three Hundred} Project \cite{300}.

\indent This paper is structured as follows: Section \ref{database} gives a brief overview of the observational catalog (SZ Large Program catalog) along with the simulation database ({\sc The Three Hundred}). Section \ref{ts_sec} describes how the simulation database has been used to build twin samples of the SZ Large Program and mentions the associated data products. Section \ref{scope} summarises the twin samples and discusses how their data products may be utilised to perform analyses that support observational studies.

\section{Overview of the observational and simulated database}
\label{database}

\subsection{Observations: The NIKA2 Sunyaev–Zeldovich Large Program}

The NIKA2 Sunyaev–Zeldovich Large Program (LPSZ) observes galaxy clusters with the NIKA2 camera \cite{n1,n2}, which is a dual--band millimeter continuum instrument installed at the IRAM 30--meter telescope located at Pico Veleta, Spain. It is capable of mapping the sky simultaneously at $150 \ \mathrm{GHz}$ and $260 \ \mathrm{GHz}$ using kinetic inductance detectors or KIDs \cite{kids}. The camera has a field of view of $6.5^{\prime}$ and a high angular resolution of $17.6^{\prime \prime}$ and $11.1^{\prime \prime}$ at  $150 \ \mathrm{GHz}$ and $260 \ \mathrm{GHz}$, respectively. The sensitivity at the two respective wavelengths is also high, standing at $ 9 \  \mathrm{and} \ 30 \ \mathrm{mJy \ s^{\frac{1}{2}}}$, respectively. Details of the performance of the camera can be found in \cite{np}. The instrument has been proven to be well--designed for mapping the ICM of galaxy clusters and their thermodynamic properties at intermediate to high redshift (see \textit{e.g.} \cite{adam2,ruppin}). 

The LPSZ is one of the four Large Programs that are a part of the NIKA2 guaranteed time. It is a high--resolution follow--up of approximately 50 SZ--selected galaxy clusters from the Planck \cite{p27} and Atacama Cosmology Telescope (ACT) \cite{hass} catalogs. The sample is representative of these clusters in mass and redshift. It spans an order of magnitude in mass with the cluster mass in the range $ 3 \leqslant M_{500} / 10^{14} \ \mathrm{M}_{\odot} \leqslant 11$ and a relatively high redshift range of $
0.5 < z < 0.9$, which is distributed over 4 equally spaced redshift bins.

The prior knowledge of the integrated SZ Compton parameters of these clusters from Planck and ACT catalogs, in combination with high--resolution data from NIKA2, allows the precise inference of the dynamical state of these clusters through their morphology and pressure profiles. The simulation data, explained in the next section, further strengthens this knowledge, creating room for improvement on current standards of ICM profile evaluation of clusters and consequently their mass estimation, in this redshift range. 
\subsection{Simulations: The Three Hundred}

To completely exploit the high--quality observational data from NIKA2, this project picks complimentary data from {\sc The Three Hundred} Project \cite{300}. Few examples of detailed analyses of various aspects of the simulated clusters in {\sc The Three Hundred} catalog can be found in \cite{3002, haggar, luca}.

These simulations are performed for a set of 324 Lagrangian regions of radius $15 \ h^{-1}{\mathrm {Mpc}}$, each centred on a galaxy cluster with $M_{500} > 4.6\times 10^{14} \ h^{-1} {\mathrm M_{\odot}}$. These clusters are the ones originally recognised as the most massive galaxy clusters (virial mass $M > 8\times 10^{14} \ h^{-1}{\mathrm M_{\odot}}$) identified at $z=0$ in the \textit{MultiDarkPlanck2} box of the simulation \textit{MultiDark}\footnote{Publically available at: https://www.cosmosim.org/} \cite{kyplin}. This is a dark--matter (DM)--only simulation and the \textit{MultiDarkPlanck2} box is a periodically defined cube with size $\sim 1.5 \ \mathrm{Gpc}$ comoving length. This cube contains $3840^3$ DM particles, each of mass $1.5\times 10^{9} \ h^{-1}{\mathrm M_{\odot}}$. As suggested by its name, this simulation assumes a cosmology based on the best--fit model and cosmological parameter estimates from the Planck 2015 cosmological results \cite{pcosmo}. 

All the regions are re-simulated at high resolution in $128$ snapshots between redshifts $z=0$ and $z=17$ using three hydrodynamical codes: \texttt{GADGETMUSIC} \cite{ymmusic}, \texttt{GADGET--X} \cite{gadgetx}, and \texttt{GIZMO-SIMBA} \cite{gizmo}. The first of the three is a standard smooth--particle--hydrodynamic (SPH) code while the other two are modern SPH codes. Apart from hydrodynamic simulations with astrophysical details, all the regions are simulated with a semi--analytical model or SAM using the codes \texttt{Galacticus}, \texttt{SAG}, and \texttt{SAGE}. The technical details such as initial conditions and box sizes related to these codes can be found in \cite{300}. {\sc The Three Hundred} Project also provides multi--wavelength information of the clusters at the centre of each of the 324 regions, and their environment. These include mock X--ray, optical, gravitational lensing, radio, and SZ Compton parameter maps of each cluster. Clearly, this simulation provides a very detailed, all--rounded picture of clusters and as such, is a mass--complete, volume--limited sample for objects with $M_{500} > 4.6\times 10^{14} \ h^{-1}{\mathrm M_{\odot}}$, over a wide range of redshift.

\section{The twin samples}
\label{ts_sec}

\begin{figure*}
\centering
\includegraphics[width=0.387\textwidth]{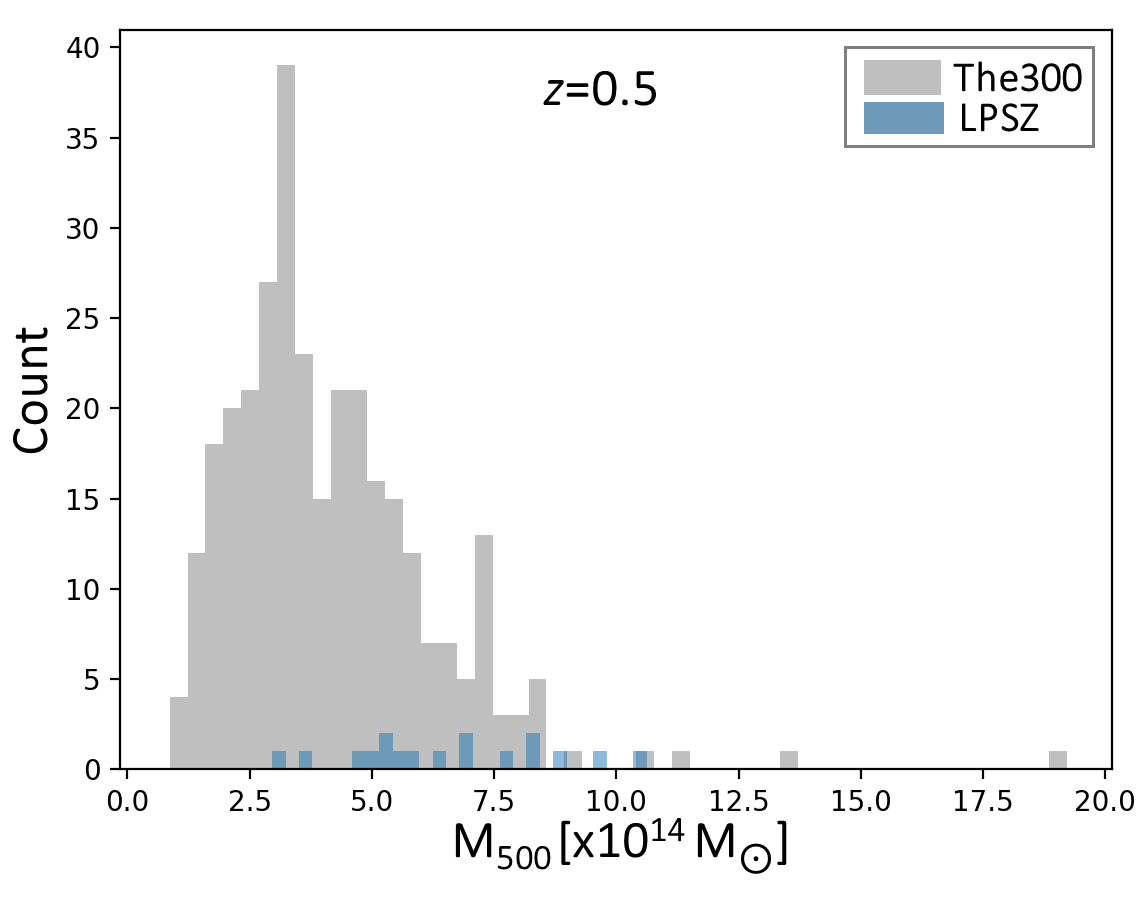}
\includegraphics[width=0.4\textwidth]{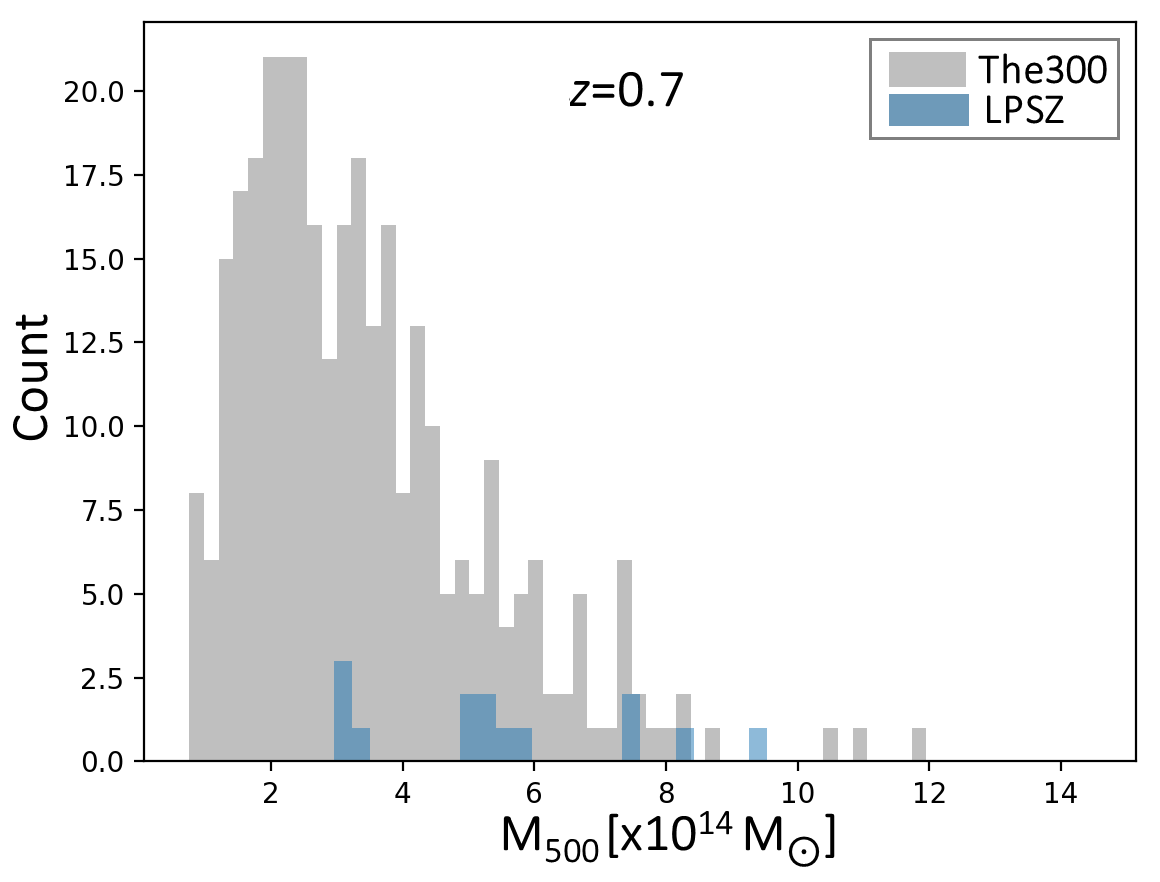}
\caption{Redshift distribution of target clusters in two of the four LPSZ redshift bins, compared to the redshift distribution of clusters in the closest identified redshift snapshot of {\sc The Three Hundred} Project.}
\label{distrired}
\end{figure*}

Some of the simulated clusters in {\sc The Three Hundred} Project have been hand--picked to generate the so--called twin samples of the LPSZ. This has been done by cross--matching different parameters of the clusters in LPSZ and {\sc The Three Hundred} Project. At first, redshift snapshots were identified in the simulation such that the median redshift of each LPSZ redshift bin is the closest to the redshift of the selected snapshot. Samples of the redshift distribution of clusters in both catalogs can be seen in Fig. \ref{distrired}. Evidently, the simulation offers a wide range of clusters to create twin samples at each of these redshifts.

\begin{figure}
\centering
\includegraphics[width=0.32\textwidth]{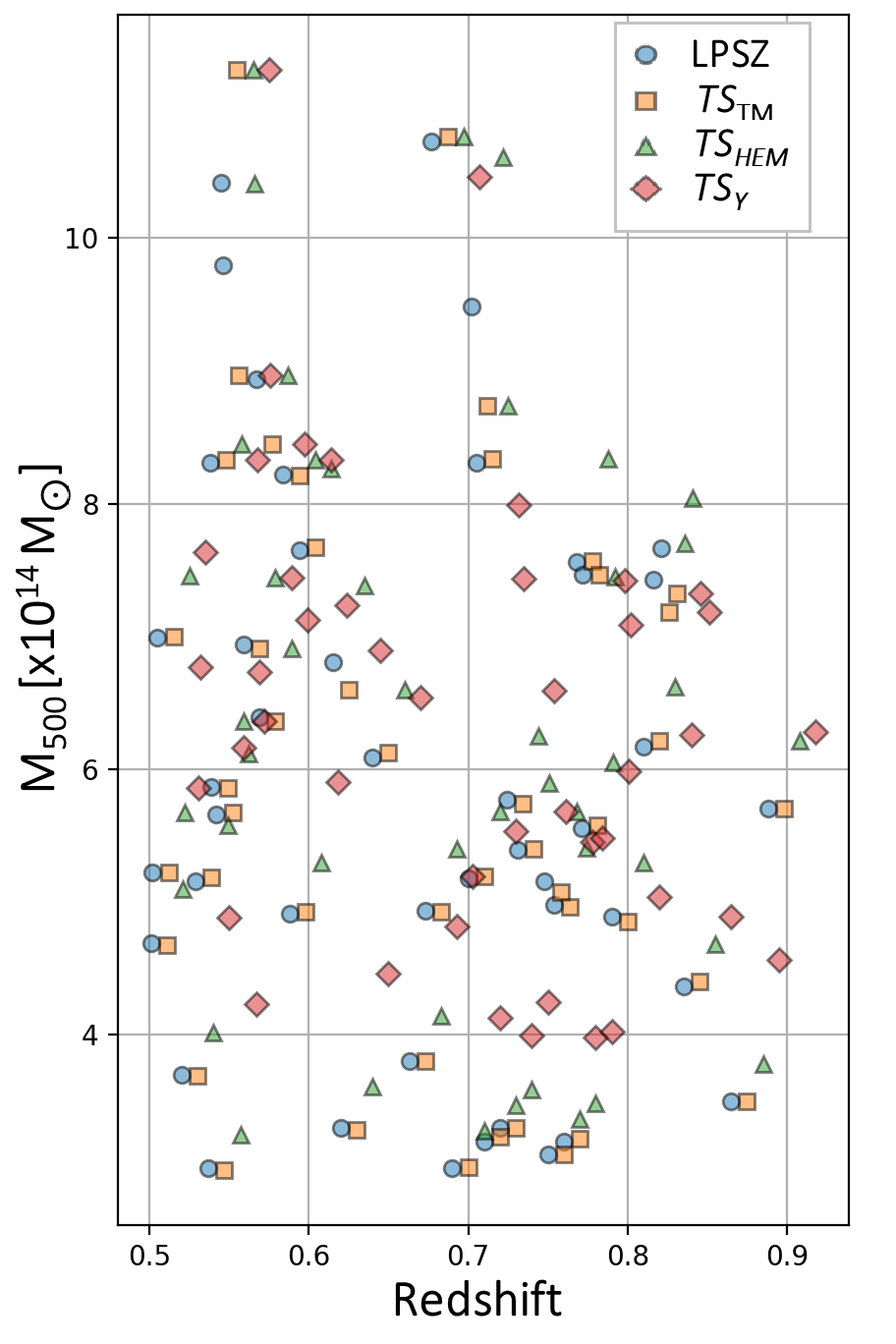} 
\caption{Redshift--mass distribution of the target LPSZ clusters compared to the twin samples generated using {\sc The Three Hundred} Project. Each twin sample uniquely represents the LPSZ catalog.}
\label{distri}
\end{figure}

After identifying simulated clusters at optimal redshifts, their following properties were cross--matched with the LPSZ clusters to, so far, generate 3 twin samples (TS): total mass ($TS_{\rm{TM}}$), hydrodynamic mass ($TS_{\rm{HEM}}$), and integrated Compton parameter at $R_{500}$ ($TS_{{Y}}$). Each of these has generated a unique sample of clusters that is, in the given criteria, equivalent to the LPSZ. This uniqueness can be visualised through the redshift--mass distribution of the twin samples compared to the LPSZ target clusters, shown in Fig. \ref{distri}. 

\begin{figure*}
\centering
 \includegraphics[width=0.41\textwidth]{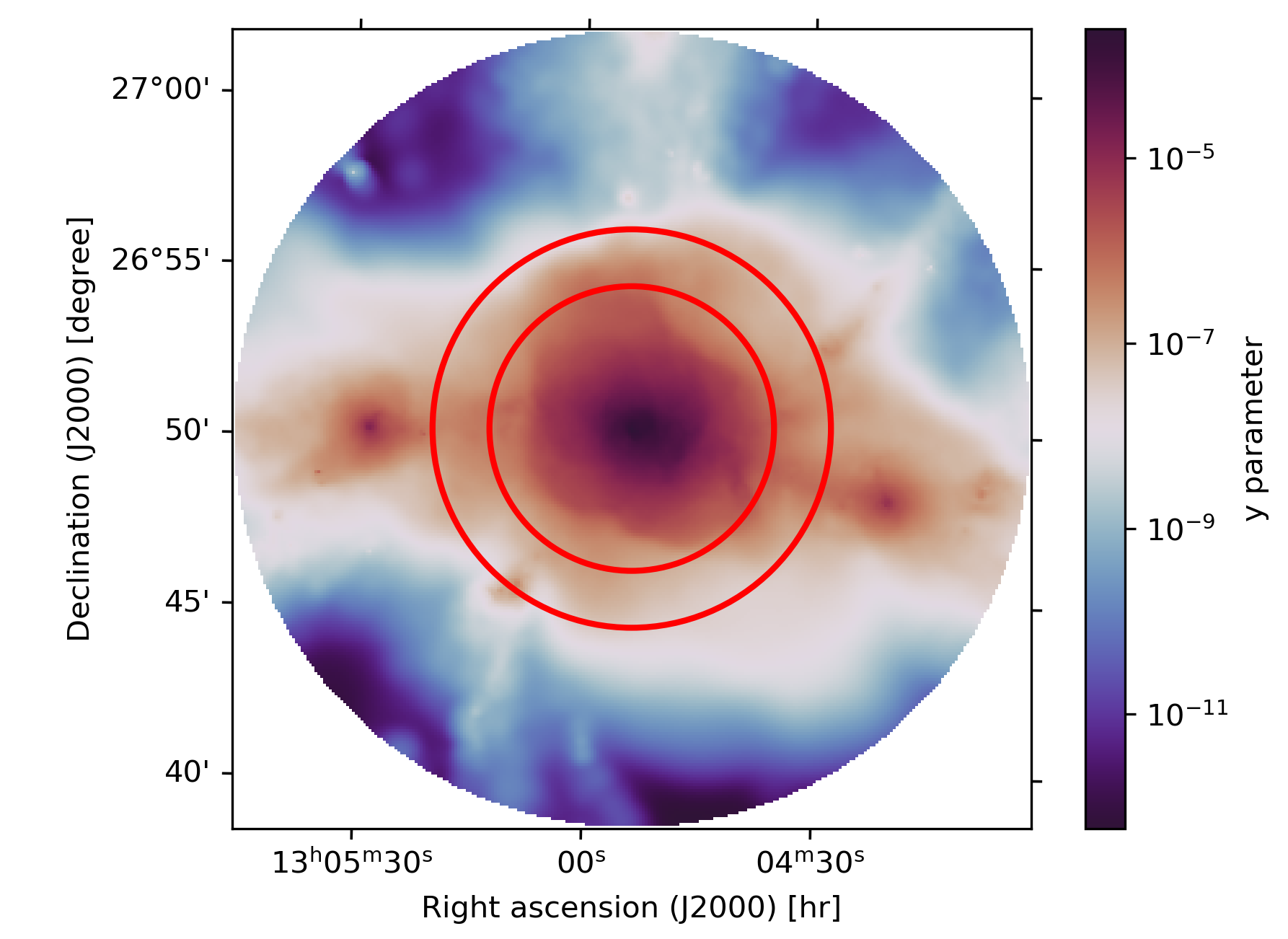}
 \includegraphics[width=0.41\textwidth]{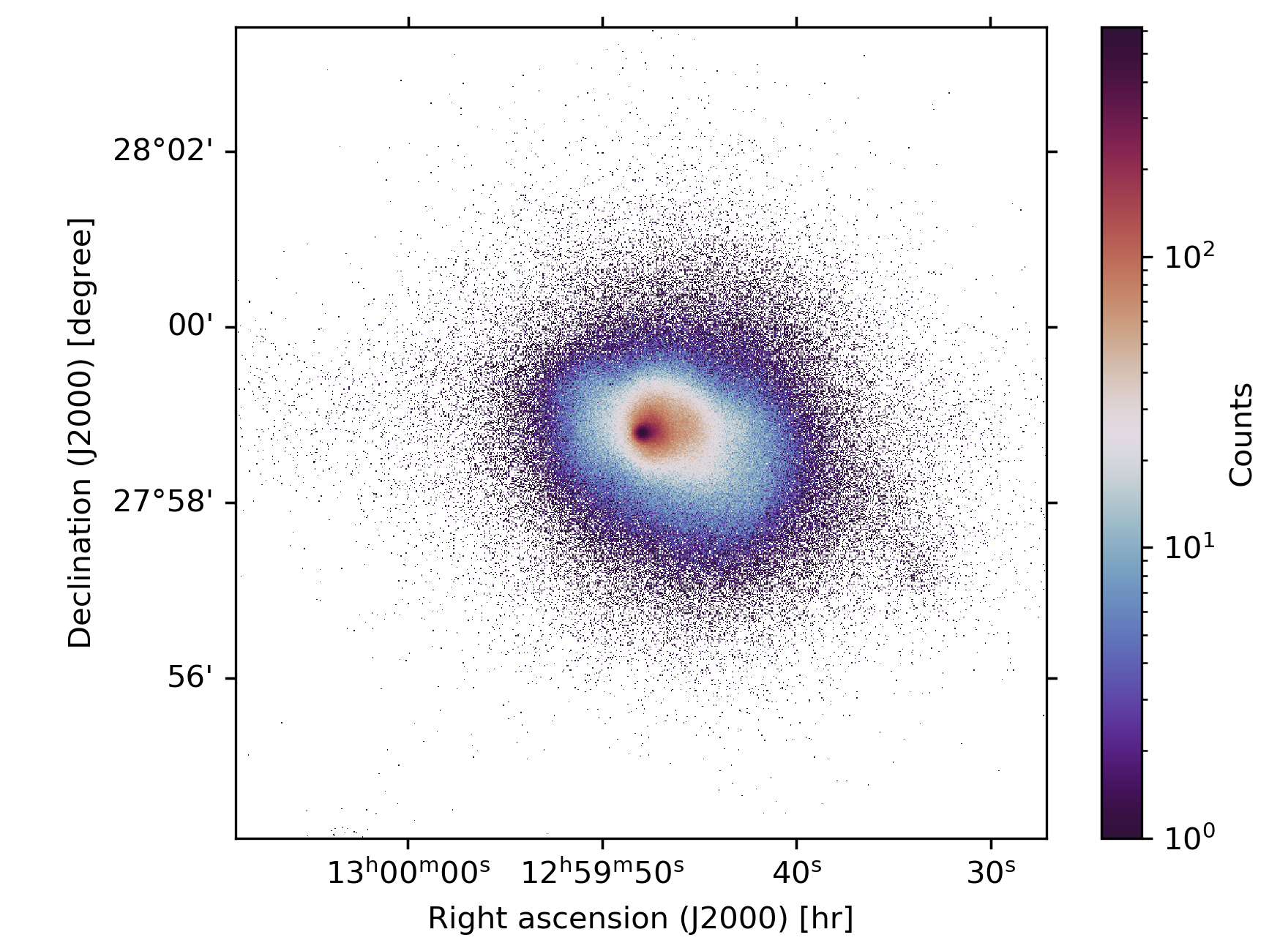}
 
\caption{Multi--wavelength maps of two clusters in {\sc The Three Hundred} simulation database. \textit{Left:} Compton parameter map for a cluster with $\log [M_{500}/ h^{-1} \ {\rm M_{\odot}] = 14.661 }$, $R_{500} = 1229.5 \ h^{-1} {\rm kpc}$, at $z=0.817$. Map size: $\sim 23.2^{\prime} \times 23.3^{\prime}$. Pixel size: $\sim 5^{\prime \prime}$. The two concentric circles represent $R_{500}$ and $R_{200}$, respectively. \textit{Right:} Mock X-Ray photon count image of a cluster with $\log  [M_{200}/ h^{-1} {\rm M_{\odot}] = 14.824 }$, $R_{200} = 1693.4 \ h^{-1} {\rm kpc}$, at $z=0.82$. Map size: $\sim 9.2^{\prime} \times 9.2^{\prime}$. Pixel size: $\sim 0.7^{\prime \prime}$.}
\label{multi}   
\end{figure*}

The average ratio of masses for $TS_{\rm{TM}}$ is $\left< M_{\rm{LPSZ}}/{M_{TS_{\rm{TM}}}}\right>=1.01 \pm 0.02$, indicating the availability of twins for each cluster in the LPSZ. On the other hand, the mass bias for the other two samples is estimated to be $\left< {M_{\rm{LPSZ}}}/{M_{TS_{\rm{HEM}}}}\right>=0.93\pm 0.02$ and $\left< {M_{\rm{LPSZ}}}/{M_{TS_{Y}}}\right>=0.9 \pm 0.1$, respectively. These values are consistent with mass bias values estimated in \cite{giulia}. The matching criterion for the integrated Compton parameter is also satisfied for $TS_{{Y}}$ with $\left<{Y_{500_{\rm{LPSZ}}}}/{Y_{500_{TS_Y}}}\right>=1.01 \pm 0.03$. As mentioned in the previous section, {\sc The Three Hundred} Project provides a multitude of multi--wavelength data products for each cluster, some of which are also provided along different lines--of--sight (see Section \ref{scope}). An example of the $Y$ map and X--ray photon count map of two clusters can be seen in Fig. \ref{multi}.

\section{Summary and outlook}
\label{scope}
Twin samples have successfully been generated from {\sc The Three Hundred} project by identifying equivalent clusters that accurately represent the NIKA2 LPSZ clusters, based on different selection criteria discussed in Section \ref{ts_sec}. These twin samples serve well for a wide range of analyses that support and test observational inferences. Below, we briefly discuss some of these undertaken analyses.

The Compton parameter maps from the simulations will be utilised to calculate $Y_{500}$ by directly using the projected data, to be compared to the same quantity derived from de--projected pressure profiles. Calculations will be done along different lines--of--sight in a given mass and redshift range to assess the dynamical state of clusters and its impact on the estimation of the ICM profiles of the cluster. These estimated values will also be used to calibrate the $Y_{500} - M_{\rm{500}}$ and $Y_{500} - M_{\rm{HE}}$ scaling relations in the LPSZ redshift range, extending the laws to higher redshifts than existing standards. Additionally, to monopolise on the high angular resolution of the NIKA2 camera, the maps will be injected with artificial point sources of varying positions and fluxes to estimate a threshold S/N and position of a point source that needs to be considered in estimating the ICM profile of the cluster, without bias.

The use of the above ICM profiles will be extended to infer the impact of redshift and morphology on mass estimation, using the X--ray maps available in the simulations. Further, the X--ray maps will be used to calibrate the $L_{\rm{X}} - M_{\rm{500}}$ and $L_{\rm{X}} - M_{\rm{HE}}$ scaling laws. Eventually, these scaling laws will be cross--correlated to the SZ scaling laws to minimise the scatter on inferred mass. Parallel use of the optical and weak lensing maps will be done for inferring observational estimates of the hydrostatic mass bias. These twin samples will also be used for cosmological analyses such as forecasting gas fraction measurements and the inference of the impact of systematics on the Hubble Constant ($H_{\rm{0}}$) measurements. Clearly, these twin samples are viable, potent tools to extend and test inferences from high--quality observations.

\end{document}